\documentclass[10pt,journal,comsoc]{IEEEtran}

\usepackage[normalem]{ulem}
\usepackage{multirow}
\usepackage{bbold}
\usepackage{ifpdf}
\usepackage{ulem} 
\usepackage{caption}
\usepackage{subcaption}%
\usepackage[cmex10]{amsmath}
\usepackage{amssymb}
\usepackage{verbatim}
\usepackage{algorithm}
\usepackage{algorithmic}
\usepackage{array}
\usepackage{latexsym}
\usepackage{color}
\usepackage{textgreek}
\usepackage{subscript}
\usepackage{rotating}
\usepackage{booktabs,multirow}
\usepackage{mdframed}	
\usepackage{tabularx}
\usepackage{amsmath}
\usepackage{makecell}
\usepackage{tabto}
\usepackage{mathrsfs}
\usepackage{url}
\usepackage{cite}
\usepackage{listings}
\usepackage{array}
\usepackage[table]{xcolor}
\usepackage{ragged2e}
\usepackage{booktabs}
\usepackage{enumitem}
\usepackage{balance}
\usepackage{comment}

\usepackage{graphicx}
\usepackage{comment}
\usepackage[cmex10]{amsmath}
\usepackage{amssymb}
\usepackage{bbold}
\usepackage{float}
\usepackage[cmex10]{amsmath}
\usepackage{amssymb}
\usepackage{verbatim}
\usepackage{array}
\usepackage{latexsym}
\usepackage{color}
\usepackage{tabularx} 
\usepackage{textgreek}
\usepackage{subscript}
\usepackage{rotating}
\usepackage{booktabs,multirow}
\usepackage{mdframed}	
\usepackage{tabularx}
\usepackage{amsmath}
\usepackage{makecell}
\usepackage{tabto}
\usepackage{mathrsfs}
\usepackage{url}
\usepackage{cite}
\usepackage{listings}
\usepackage{array}
\usepackage[table]{xcolor}

\definecolor{pinkpurple}{rgb}{0.6, 0.1, 0.9} % A mix of pink and purple
\begin{document}

\title{Stochastic Modeling of Human-Machine Authentication Channels under Partial Information Leakage}
% === AUTHORS ===
\author{
\IEEEauthorblockN{Nilesh Chakraborty~\IEEEmembership{Member IEEE}, Mohammad Zulkernine}~\IEEEmembership{SMIEEE}, Burak Kantarci~\IEEEmembership{SMIEEE}\\
\thanks{N. Charkraborty and Burak Kantarci are with the University of Ottawa, Ottawa, ON, Canada. Emails: \{nchakrab,burak.kantarci\}@uottawa.ca\\
 M. Zulkernine is with the School of Computing, Queen's University, Kingston, ON, Canada. Email: mz@queensu.ca

}\vspace{-0.3in}
}

% === HEADER ===
%\markboth{IEEE Transactions on Network and Service Management}%{Shell \MakeLowercase{\textit{et al.}}: LM-based Automated Generation of Linux TC Configurations}

\IEEEtitleabstractindextext{
% === ABSTRACT ===
\begin{abstract}
Reliable and secure human-machine communication is fundamental to IoT and cyber-physical ecosystems, where smartphones and wearables 
commonly serve as authentication controllers. PIN-based authentication can be viewed as a low-bandwidth communication channel through 
which users transmit numeric credentials under practical constraints. However, conventional evaluations adopt a binary view of 
security$-$treating such channels as either fully secure or fully compromised$-$thereby overlooking the progressive reliability 
degradation caused by partial information leakage in real-world IoT settings.
In this paper, we model the PIN entry process as a stochastic human-IoT communication system and propose a context-conditioned 
probabilistic inference framework to quantify reliability loss and Quality-of-Service (QoS) degradation under partial symbol exposure. 
The proposed approach treats missing digits as latent variables and estimates them using smoothed conditional probability distributions 
with fallback priors. Unlike traditional sequential models that assume contiguous positional dependencies, the method does not 
explicitly parameterize hidden-state transitions or emissions; instead, it performs context-driven probabilistic inference to 
approximate latent dependencies across digit positions.
Using over one million real-world four-digit PIN samples, we evaluate single-, double-, and triple-digit leakage 
scenarios and derive position-dependent reliability metrics. The proposed model achieves up to 55.31\% prediction 
accuracy for one missing digit and 12.12\% for three missing digits, while consistently outperforming a standard 
sequence-model baseline and classical machine learning models in terms of precision, recall, and F1-score. These 
results formalize PIN entry as a noisy human--IoT communication channel and demonstrate substantial reliability 
degradation under realistic partial exposure conditions.
\end{abstract}

% === KEYWORDS ===
}

\pagestyle{empty}

\maketitle
\thispagestyle{empty}
\IEEEdisplaynontitleabstractindextext
\IEEEpeerreviewmaketitle

\section{Introduction}
The reliability of communication processes involving human interaction is a fundamental aspect of modern intelligent networks, cyber-physical systems, and IoT-based infrastructures~\cite{singh2024reliability, lee2024minding, reliability-communication2015}. Among these, \textit{PIN-based authentication} exemplifies a low-bandwidth human-machine communication channel where the user transmits a short numeric message for identity verification. Such channels are inherently vulnerable to partial information leakage caused by environmental factors (e.g., camera placement, lighting), cognitive constraints (e.g., typing speed, attention), or side channels in the physical communication medium (e.g., wireless signal perturbations)~\cite{modern_shoulderSurfing, bace2022privacyscout, tan2023hollow}. Conventional authentication models simplify the process by adopting a binary notion of success$:$ a PIN is viewed as either entirely secure or completely compromised. Such simplification neglects the intermediate reliability losses that arise under partial observation.

\begin{figure}[!ht]
    \centering
    \includegraphics[width=\linewidth]{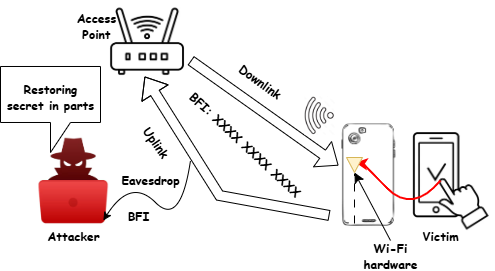}
    \caption{Eavesdropping clear-text BFI transmitted to the access point, enabling secret information leakage.}
    \label{fig:csi-attack}
\end{figure}

Recent studies have shown that human-centric and physical-layer interactions introduce observable side-channel patterns in communication systems~\cite{hu2023password, kwon-shoulderSurfing, chakraborty2025your}. In particular, Fig.~\ref{fig:csi-attack} illustrates how Channel State Information (CSI) feedback, transmitted in clear text during Wi-Fi communication, can unintentionally act as a leakage pathway. Here, the finger movements of a user entering a PIN modulate the wireless channel, creating measurable distortions in the Beamforming Feedback Information (BFI)~\cite{hu2023password}. This scenario can be interpreted as a form of QoS degradation in the human-machine communication link, where unintentional signal perturbations convey partial message content to an unintended receiver. In this work, we interpret such QoS degradation in terms of the recoverability of the transmitted PIN under partial observability, which we quantify through inference performance metrics (e.g., accuracy, precision, recall, and F1-score) from the perspective of an observing entity. Under this interpretation, partial leakage events are analogous to symbol errors or reliability losses in conventional communication systems, where the integrity of message delivery must be assessed probabilistically rather than deterministically.\\

\noindent
To systematically model the above leakage, we contextualize it within the well-documented Human-Centric Shoulder-Surfing (HCSS) threat in IoT environments~\cite{nileshIoT-PIN, IoTSec-PIN2018}. This class of attacks is known to facilitate such information exposure~\cite{kwon-shoulderSurfing,chakraborty2014improved,roth2004pin}.
While prior work analyzes partial PIN compromise with a single missing digit~\cite{chakraborty2025your}, real-world observations yield more varied incomplete views shaped by vantage point, device UI, and user behavior. Based on these observations, we make the following contributions. 

\begin{itemize}
    \item Motivated by the interpretability and data efficiency of probabilistic sequence modeling approaches, we propose an HMM-inspired reliability estimator that captures contextual dependencies among PIN digits and quantifies residual communication reliability under partial exposure$-$a noisy human-IoT interaction channel susceptible to side-channel leakage.

    \item Using more than one million real-world PIN samples derived from the RockYou dataset~\cite{skullsecurity-passwords}, our framework$-$for the first time$-$evaluates the impact of partial observation across multiple exposure levels. The results show that traditional binary security assumptions significantly overestimate the reliability of such communication channels.
    
    \item We conduct a comprehensive comparative analysis across classical machine learning baselines$-$Decision Tree, Random Forest, and Naive Bayes$-$and include a standard HMM-based sequence model as a reference. The results show that the proposed context-conditioned probabilistic model achieves higher accuracy under multi-digit leakage, establishing a practical reliability framework for quantifying how partial compromises propagate through interconnected human-machine authentication systems.
\end{itemize}

\noindent 
The rest of this paper is structured as follows. Section \ref{sec:relatedwork} reviews prior work on shoulder-surfing, human-centered adversarial models, and probabilistic authentication. Section \ref{sec:methodology} details our modeling approach, including smoothing and inference for single- and multi-digit prediction. Section \ref{sec:experimental-results} reports results, analyzing digit position effects and comparing HMM with Decision Tree, Random Forest, and Naive Bayes baselines. Section \ref{sec:conclusionFW} concludes with future research directions.

\section{Related Work}
\label{sec:relatedwork}

IoT device authentication continues to face multiple open challenges. These include the heterogeneity of device capabilities, partial information leakage via side channels, and enduring susceptibility to advanced guessing attacks.
Such limitations considerably weaken authentication reliability in practical smart environments, where mobile and wearable devices often function as primary authentication controllers~\cite{kokila2025authentication}. Existing defenses focus primarily on blocking direct observation, yet partial exposure~\cite{chakraborty2025your} and adaptive adversarial strategies~\cite{modern_shoulderSurfing} reveal substantial reliability and usability challenges that are unique to modern IoT settings.

Among the various leakage vectors, shoulder-surfing attacks have emerged as a dominant class of observational threats that compromise user input privacy across both mobile and IoT interfaces~\cite{schneegass2022investigation, nileshIoT-PIN}. These attacks have evolved from simple visual observation of keypad inputs to increasingly sophisticated modalities. Video-based techniques exploit cameras and computer vision to reconstruct input sequences remotely~\cite{shukla2019stealing}, while thermal imaging leverages residual heat signatures on keypads or screens~\cite{wodo2016thermal}. Audio-based approaches further demonstrate that keystroke sounds alone can leak authentication information, enabling inference of haptic patterns~\cite{varma2022vibroauth}. Collectively, these studies highlight the diverse range of modalities adversaries may exploit.

In parallel, a distinct line of research emphasizes HCSS, which focuses on the perceptual and cognitive abilities of unaided human attackers~\cite{kwon-shoulderSurfing}. Roth et al.~\cite{roth2004pin} introduced challenge-response protocols to exploit attackers’ cognitive limits, while Kwon et al.~\cite{kwon-shoulderSurfing} demonstrated that HCSS attackers can bypass such defenses. Subsequent work further confirmed that human observers are often more capable than traditional security analyses assume~\cite{chakraborty2014improved}.

Complementing the above human-centric studies, probabilistic models have been widely used to analyze authentication data. Hidden Markov Models (HMMs) capture sequential dependencies between characters, handle sparse data via smoothing, and yield interpretable transition probabilities~\cite{ma2014study, thai2024study}. \textit{Probabilistic Context-Free Grammars} (PCFGs) have achieved strong results in password analysis by exploiting the structural diversity of mixed character sets~\cite{weir2009password}. However, their effectiveness is limited for PINs, which consist solely of fixed-length numeric sequences. In such cases, PCFG templates collapse to trivial forms (e.g., DDDD), providing little advantage over simpler sequence models while adding computational overhead.

By contrast, probabilistic sequence models such as HMMs, n-grams, and context-conditioned frequency models remain well-suited for PIN analysis. These approaches are data-efficient, generalizing effectively even from small datasets through smoothing, robust against overfitting in low-data regimes, and interpretable, as they capture dependencies between digits in a transparent manner. In particular, context-conditioned models estimate probabilities directly from observed digit patterns, providing a practical alternative to fully parameterized generative models when the goal is inference under partial observation. 

\begin{comment}
Empirically, such context-conditioned approaches can also better capture cross-position dependencies compared to standard HMM formulations, especially in partially observed PIN inference settings. These characteristics make probabilistic approaches particularly suitable in settings where training data is limited and interpretability is important, whereas neural methods typically benefit from larger datasets and emphasize predictive performance over transparency~\cite{bonneau2012quest}.
\end{comment}

\begin{table*}[h]
\centering
\caption{Summary of Notations}
\resizebox{0.6\textwidth}{!}{
\begin{tabular}{|c|c|}
\hline
\textbf{Notation} & \textbf{Meaning} \\
\hline
$\mathbf{d}$ & Four-digit PIN represented as $[d_1, d_2, d_3, d_4]$ \\
\hline
$d_i$ & Individual digit at position $i$ in the PIN \\
\hline
$\mathcal{D}$ & Digit space $\{0, 1, \ldots, 9\}$ \\
\hline
$|\mathcal{D}|$ & Size of the digit space (equals 10) \\
\hline
$\hat{d}_i$ & Predicted value for digit at position $i$ \\
\hline
$M$ & Set of missing (unknown) digit positions, $M \subseteq \{1, 2, 3, 4\}$ \\
\hline
$O$ & Set of observed (known) digit positions, $O \subseteq \{1, 2, 3, 4\}$ \\
\hline
$|M|$ & Number of missing digits \\
\hline
$\mathbf{d}_M$ & Subsequence of digits at positions in $M$ (missing digits) \\
\hline
$\mathbf{d}_O$ & Subsequence of digits at positions in $O$ (observed digits) \\
\hline
$\mathbf{m}$ & Vector of candidate values for missing digits, $\mathbf{m} = [m_i : i \in M]$ \\
\hline
$\mathbf{o}$ & Vector of known digit values at observed positions, $\mathbf{o} = [o_j : j \in O]$ \\
\hline
$\hat{\mathbf{m}}$ & Predicted values for all missing digits \\
\hline
$C$ & Context representing the observed digits \\
\hline
$x$ & A candidate digit value \\
\hline
$P(x \mid C)$ & Conditional probability of digit $x$ given context $C$ \\
\hline
$N(x, C)$ & Frequency of digit $x$ observed in context $C$ \\
\hline
$N(C)$ & Total frequency of context $C$, computed as $\sum_{y \in \mathcal{D}} N(y, C)$ \\
\hline
$\alpha$ & Laplace smoothing parameter (set to 1.0) \\
\hline
$P_{\text{prior}}(x)$ & Global prior probability of digit $x$ \\
\hline
$N(x)$ & Overall frequency of digit $x$ in the training set \\
\hline
$N$ & Total frequency across all digits, $N = \sum_{y \in \mathcal{D}} N(y)$ \\
\hline
$(a, b)$, $(a, b, c)$, etc. & Candidate digit values for multiple missing positions \\
\hline
$[a, b, c, d]$ & Complete four-digit PIN sequence \\
\hline
$[*, *, c, d]$ & Partial PIN pattern where $*$ denotes any digit (wildcard) \\
\hline
\end{tabular}}
\label{tab:summary_notations}
\end{table*}

\section{Methodology}
\label{sec:methodology}
In an IoT setting where smartphones act as controllers for smart locks and services, user secrets are frequently entered on touch interfaces in public or semi-public spaces. This exposes the authentication process to HCSS and recording-based attacks (e.g., single or multi-observer/video), which may reveal all or part of a PIN during entry. Even partial exposure (one or more digits) can significantly reduce the search space for the full code~\cite{IoTSec-PIN2018}.
Complementing these visual leaks, non-visual RF side-channels further enable \textit{overhearing} of keystrokes. For instance, WiKI-Eve exploits clear-text Wi-Fi BFI transmitted from smartphones to access points, which can be captured by nearby devices operating in monitor mode to infer keystrokes and reconstruct secrets with high accuracy in realistic public Wi-Fi environments~\cite{hu2023password}.
Together, these findings motivate a threat model in which an attacker, positioned with either line-of-sight or RF vantage, obtains partial and uncertain observations of the PIN and subsequently performs intelligent inference to recover the full sequence~\cite{hu2023password, IoTSec-PIN2018}.

To operationalize this threat model in practical IoT side-channel settings, such as CSI/BFI-based keystroke eavesdropping, the attacker typically does not obtain a full PIN directly. Instead, the leakage process yields uncertain evidence about individual keypresses, with some digits inferred at high confidence and others remaining ambiguous or unrecoverable. Our masking formulation models this inference-level outcome rather than the underlying signal-generation process. Specifically, observed digits correspond to symbols that a side-channel classifier or observer recovers with sufficient confidence, whereas masked digits represent positions for which the evidence is noisy, weak, or conflicting. Under this abstraction, partial digit masking serves as a tractable surrogate for confidence-varying side-channel leakage in human-machine authentication channels.

Thus, we model partial PIN exposure as a probabilistic inference problem, where an adversary seeks to predict unknown digits based on observed partial information. Let $\mathbf{d} = [d_1, d_2, d_3, d_4]$ denote a four-digit PIN, where each digit $di \in \mathcal{D} = {0,1,\ldots,9}$. The objective is to estimate the most likely values of the missing digits given the known ones. In Table~\ref{tab:summary_notations}, we summarize all the notations used in this section.\\

\noindent 
\textit{Problem Formulation and Probabilistic Inference Framework:} 
For single missing digit scenarios, we seek to predict the unknown digit by maximizing the conditional probability given the observed context. Predicting the first digit when the remaining three are known is expressed as:
\[
\hat{d}_1 = \arg\max_{d \in \mathcal{D}} P(d_1 = d \mid d_2, d_3, d_4),
\]
where $\hat{d}_1$ represents the predicted value for the first digit, and the conditioning context consists of the observed digits $d_2, d_3, d_4$.

For multiple missing digits, we generalize this formulation. Let $M \subseteq \{1, 2, 3, 4\}$ denote the set of missing digit positions and $O \subseteq \{1, 2, 3, 4\}$ represent the set of observed (known) digit positions, where $O \cup M = \{1, 2, 3, 4\}$ and $O \cap M = \emptyset$. The prediction problem becomes:
\[
\hat{\mathbf{m}} = \arg\max_{\mathbf{m} \in \mathcal{D}^{|M|}} P(\mathbf{d}_M = \mathbf{m} \mid \mathbf{d}_O = \mathbf{o}),
\]
where $\mathbf{d}_M$ denotes the subsequence of digits at positions in $M$, $\mathbf{d}_O$ denotes the subsequence of digits at positions in $O$, $\mathbf{m} = [m_i : i \in M]$, and $\mathbf{o} = [o_j : j \in O]$.
For example, predicting the first two digits given knowledge of the last two requires:
\[
(\hat{d}_1, \hat{d}_2) = \arg\max_{(a,b) \in \mathcal{D}^2} P(d_1=a, d_2=b \mid d_3, d_4),
\]
where $(a,b)$ represents all possible digit pairs and the context $d_3, d_4$ provides the conditioning information.

%-------------------------

\noindent 
It is important to note that the proposed model adopts a context-conditioned probabilistic formulation, where the likelihood of missing digits is estimated directly from observed positional patterns in the data. For a given context $C$, the probability of a missing digit $x$ is estimated using maximum likelihood estimation with smoothing and fallback priors to ensure robustness under sparse observations.

While inspired by sequence modeling principles, this formulation does not explicitly parameterize hidden-state transitions and emissions as in classical HMMs. Instead, it performs inference through context-dependent conditional distributions, enabling effective modeling of cross-position dependencies in partially observed PIN sequences.

%-------------------------

\noindent 
\textit{Laplace Smoothing and Global Prior:}  
To mitigate sparsity and avoid zero-probability estimates for unseen events, we apply Laplace (add-$\alpha$) smoothing with $\alpha=1.0$:
\[
P(x \mid C) = \frac{N(x,C) + \alpha}{N(C) + \alpha \cdot |\mathcal{D}|},
\]
where $N(x,C)$ is the frequency of digit $x$ observed in context $C$, $N(C) = \sum_{y \in \mathcal{D}} N(y,C)$ is the total frequency of $C$, and $|\mathcal{D}|=10$.

For unseen contexts, the model falls back to a global prior:
\[
P(x \mid C) \approx P_{\text{prior}}(x) = 
\frac{N(x)+\alpha}{N + \alpha \cdot |\mathcal{D}|},
\]
where $N(x)$ is the overall frequency of digit $x$ and $N=\sum_{y \in \mathcal{D}} N(y)$.

%-------------------------

\noindent 
\textit{Inference and Prediction:} Missing digits are predicted using the learned conditional distributions.

\noindent
\textit{Single missing digit:}
\[
\hat{d} = \arg\max_{x \in \mathcal{D}} P(x \mid C).
\]

\noindent
\textit{Two missing digits:}
Let \( C = (d_3, d_4) \). If sufficient training samples exist, we estimate the joint distribution:
\[
P(d_1=a,d_2=b \mid C) =
\frac{N([a,b,C])+\alpha}{N([*,*,C])+\alpha \cdot 10^2}.
\]

For sparse contexts, we approximate:
\[
P(d_1=a,d_2=b \mid C)
\approx P(d_1=a \mid C)\times P(d_2=b \mid C).
\]

\noindent
\textit{Three missing digits:}
Let \( C = (d_4) \). Then:
\[
\begin{aligned}
P(d_1=a,d_2=b,d_3=c \mid C)
&\approx P(d_1=a \mid C) \\
&\quad \times P(d_2=b \mid C) \\
&\quad \times P(d_3=c \mid C).
\end{aligned}
\]

We employ direct joint estimation only when the conditioning context occurs at least $\tau=10$ times in training; otherwise, the independence approximation is used. To justify this choice, we conducted a sensitivity analysis on $(d_1,d_2 \mid d_3,d_4)$ with $\tau \in \{1,5,10,20,50\}$. The prediction accuracy remained stable at 31.83\% with a 95\% confidence interval of [31.67\%, 32.00\%] across all values, indicating that the model is not sensitive to $\tau$ in sufficiently dense contexts and supporting $\tau=10$ as a robust threshold.

\section{Results}
\label{sec:experimental-results}
\noindent
\textit{Dataset Preparation:} Following the dataset derivation approach of Wang et al.~\cite{wang2017understanding} and our prior work~\cite{chakraborty2025your}, we extracted over one million four-digit PINs from the RockYou password leak. We retained numeric substrings that exhibited precisely four consecutive digits while filtering out longer or alphanumeric strings to preserve behavioral fidelity. We employ an $80$-$20$ train-test split with random state $39$ to ensure reproducibility.\\

\noindent  
\textit{Training Process:} The model training involves extracting digit-context frequency counts from the training dataset. For each PIN $\mathbf{d} = [d_1, d_2, d_3, d_4]$, we collect statistics for all possible partial exposure scenarios. For instance, to train the first-digit predictor, we extract tuples $(d_1; d_2, d_3, d_4)$ where $d_1$ serves as the target and $d_2, d_3, d_4$ form the context. Similar extraction processes generate training data for all other single-digit, two-digit, and three-digit prediction scenarios.\\

\begin{figure}
    \centering
    \includegraphics[width=0.48\textwidth]{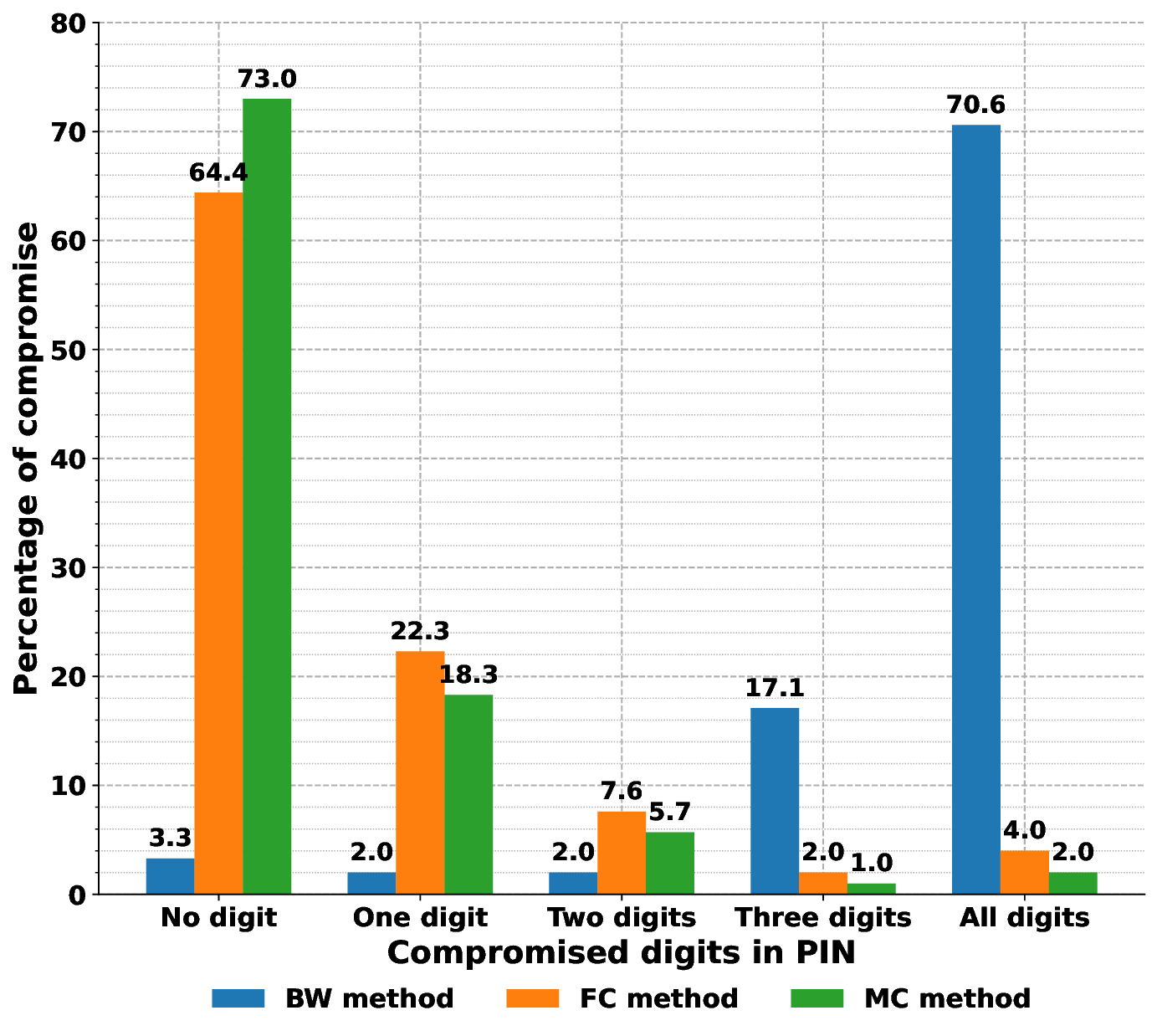}
    \caption{Performance of trained shoulder surfers with probabilistic inference support in revealing PIN digits under three defense mechanisms: the basic BW method~\cite{roth2004pin}, the advanced FC method~\cite{kwon-shoulderSurfing}, and the lately developed MC method~\cite{chakraborty2014improved}}
    \label{fig:problemstatementRevised}
\end{figure}

\begin{figure}[!ht]
    \centering
    \includegraphics[width=\linewidth]{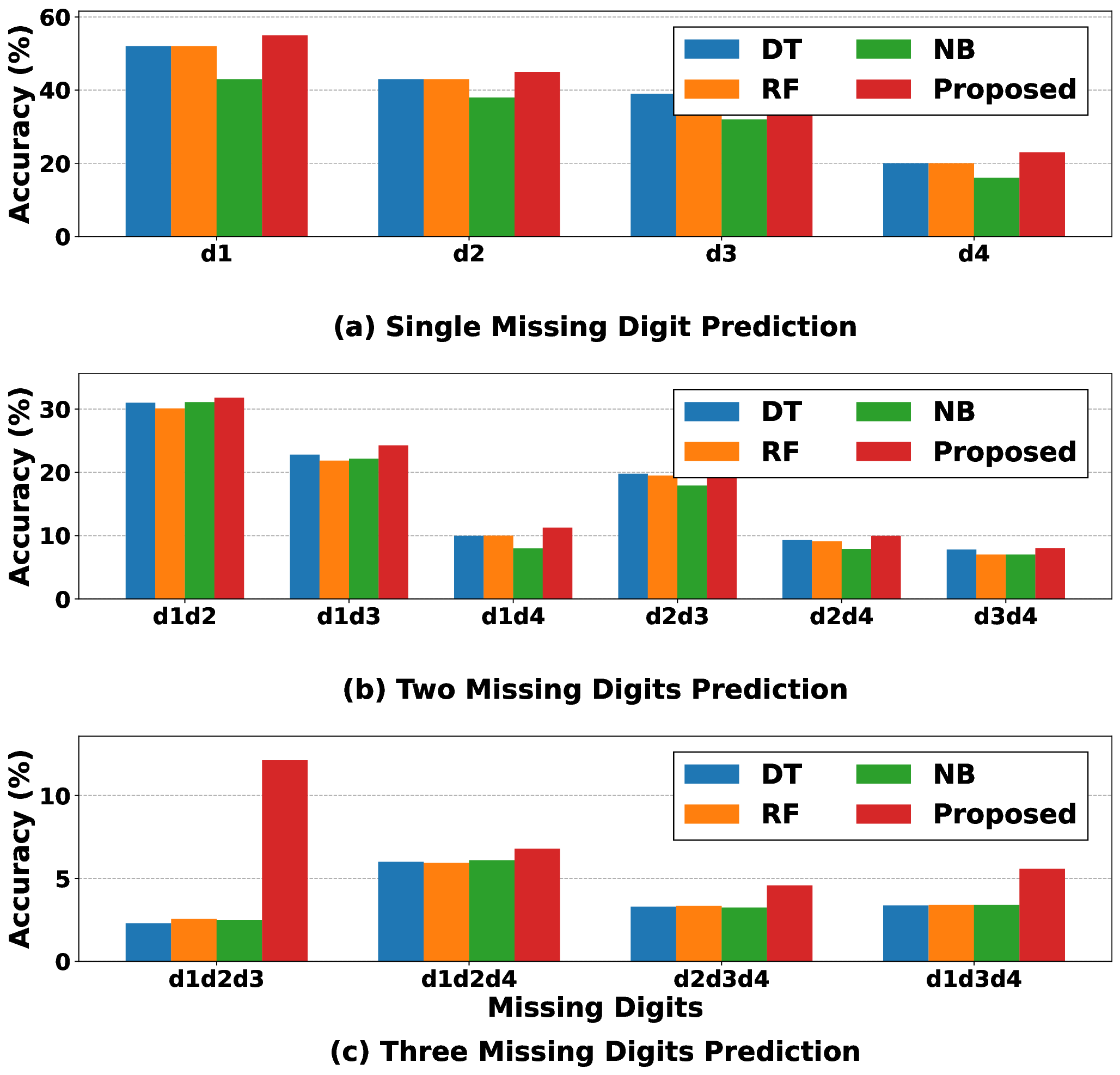}
    \caption{Comparison of prediction accuracy across Decision Tree (DT), Random Forest (RF), Naive Bayes (NB), and the proposed model under single, double, and triple-digit missing scenarios.}
    \label{fig:comp_prediction}
\end{figure}

\noindent
\textit{Experimental Design:} We systematically evaluate all possible partial exposure scenarios: four single-digit predictions ($d_i$ given $\{d_j : j \neq i\}$), six two-digit predictions (all $\binom{4}{2}$ combinations), and four three-digit predictions (each individual digit serving as the sole observed information). For each scenario, we compute accuracy (fraction of correct predictions), macro-averaged precision, recall, and F1-score across all digit classes. The HMM model is included as a baseline to evaluate the effectiveness of the proposed context-conditioned probabilistic approach.\\

\noindent 
\textit{Single Digit Prediction Analysis:} The upper portion of Table~\ref{tab:missing_digitsPred} summarizes single missing digit results. Prediction accuracy declines monotonically from the first to the fourth digit, reflecting positional dependencies in human PIN selection. The first digit achieves the highest accuracy ($55.31\%$), revealing strong prefix regularities, whereas the fourth digit drops to $23.62\%$, yet remains over twice the random baseline.

The lower portion of Table~\ref{tab:missing_digitsPred} reports accuracy under multi-digit loss ($2$ and $3$ missing digits). Despite increased uncertainty, results remain substantially above random expectations: $31.79\%$ for $d_1d_2$ (vs.\ $1\%$ random) and $12.12\%$ for $d_1d_2d_3$ (vs.\ $0.1\%$ random). These outcomes confirm that positional patterns persist even under partial exposure and further demonstrate that context-conditioned probabilistic modeling effectively captures these dependencies beyond classical sequence-based baselines.

Per-digit recall analysis further shows a strong bias toward specific prefixes$-$digits $1$, $2$, and $9$ exhibit the highest recall ($83\%$, $56\%$, and $55\%$), while $3$, $7$, and $8$ remain least predictable ($6$–$10\%$)$-$highlighting uneven structural regularities in user-generated PINs.

\begin{table}[htbp]
\caption{Missing Digits Prediction Performance. The highlighted cells indicate the highest accuracy achieved in a category.}
\begin{center}
\begin{tabular}{|c|c|c|c|c|c|}
\hline
\textbf{Target} & \textbf{Known} & \textbf{Accuracy} & \textbf{Precision} & \textbf{F1-Score} & \textbf{Recall} \\
\hline
\multicolumn{6}{|c|}{\cellcolor{gray!25}\textbf{Single Missing Digit}} \\
\hline
d1 & d2d3d4 &\cellcolor{green!25} 55.31\% & 37.57\% & 25.74\% & 20.06\% \\
\hline
d2 & d1d3d4 & 45.19\% & 31.00\% & 29.95\% & 29.00\% \\
\hline
d3 & d1d2d4 & 39.07\% & 36.70\% & 30.28\% & 25.20\% \\
\hline
d4 & d1d2d3 & 23.62\% & 20.40\% & 19.45\% & 18.70\% \\
\hline
\hline
\multicolumn{6}{|c|}{\cellcolor{gray!25}\textbf{Two Missing Digits}} \\
\hline
d1d2 & d3d4 & \cellcolor{green!25}31.79\% & 2.43\% & 2.83\% & 3.40\% \\
\hline
d1d3 & d2d4 & 24.77\% & 3.77\% & 4.18\% & 4.70\% \\
\hline
d1d4 & d2d3 & 11.27\% & 4.37\% & 4.51\% & 4.70\% \\
\hline
d2d3 & d1d4 & 21.27\% & 4.91\% & 4.92\% & 4.95\% \\
\hline
d2d4 & d1d3 & 9.98\% & 4.32\% & 4.04\% & 3.80\% \\
\hline
d3d4 & d1d2 & 9.04\% & 5.04\% & 4.80\% & 4.60\% \\
\hline\hline
\multicolumn{6}{|c|}{\cellcolor{gray!25}\textbf{Three Missing Digits}} \\
\hline
d2d3d4 & d1 & 4.58\% & 0.03\% & 0.06\% & 0.04\% \\
\hline
d1d3d4 & d2 & 5.58\% & 0.04\% & 0.08\% & 0.06\% \\
\hline
d1d2d4 & d3 & 6.79\% & 0.06\% & 0.10\% & 0.08\% \\
\hline
d1d2d3 & d4 & \cellcolor{green!25}12.12\% & 0.05\% & 0.08\% & 0.07\% \\
\hline
\end{tabular}
\label{tab:missing_digitsPred}
\end{center}
\end{table}

%-------------------
\noindent 
Building on the findings above, we deploy the proposed probabilistic inference model to assess its impact on attacker performance reported in~\cite{chakraborty2014improved}. By leveraging successfully revealed digits to predict the remaining ones, we observe noticeable improvements in attack success rates, as shown in Fig.~\ref{fig:problemstatementRevised}. 

Previous results indicate that without probabilistic inference, attackers already achieve non-trivial success in partial compromises; for example, $26.6\%$ of attacks on the FC method reveal all but one digit, and $14.6\%$ of attacks on the BW method leave only a single digit unrevealed~\cite{chakraborty2014improved, kwon-shoulderSurfing}. With probabilistic inference support, however, many of these partial compromises are transformed into full recovery: complete PIN exposure increases from $68.3\%$ to $70.6\%$ for the BW method, from $0\%$ to $4\%$ for the FC method, and from $0\%$ to $2\%$ for the MC method. This shift demonstrates that context-conditioned inference enables adversaries to exploit partial leakage more effectively.

\smallskip
\noindent\textit{Comparative analysis.}
To further evaluate the effectiveness of the proposed model, we compare it against a sequence-aware bigram baseline, which represents the simplest form of an n-gram (order-1 Markov) model capturing adjacent digit dependencies. Unlike the proposed joint model, which estimates $P(d_1,d_3 \mid d_2,d_4)$, the bigram baseline approximates this as $P(d_1 \mid d_2)P(d_3 \mid d_4)$, thereby assuming conditional independence between the two missing digits and ignoring cross-position dependencies. While higher-order n-gram or Markov models could capture longer-range dependencies, the bigram model provides a conservative and interpretable baseline for evaluating the impact of local sequence structure.

We note that the choice of the $(d_1,d_3 \mid d_2,d_4)$ configuration is made \textit{without loss of generality}. The same formulation applies to all two-digit inference scenarios, and the observed performance trends are consistent across these cases. This instance is presented as a representative example for clarity and conciseness.

\begin{table}[t]
\centering
\caption{Comparison of the proposed context-conditioned probabilistic model with a bigram baseline for predicting $(d_1,d_3)$ from $(d_2,d_4)$. Top-$k$ success rates correspond to the probability of successful PIN recovery within $k$ attempts under rate-limited attack settings.}
\label{tab:bigram_comparison}
\resizebox{0.48\textwidth}{!}{
\begin{tabular}{|l|c|c|c|c|c|}
\hline
\textbf{Model} & \textbf{Top-1} & \textbf{Top-3} & \textbf{Top-5} & \textbf{Top-10} & \textbf{Exp. Rank} \\
\hline
Proposed  & 0.2477 & 0.4187 & 0.5038 & 0.6215 & 17.75 \\
Bigram                         & 0.0954 & 0.2056 & 0.3241 & 0.4945 & 23.50 \\
\hline
\end{tabular}}
\end{table}

Table~\ref{tab:bigram_comparison} reports both prediction accuracy and rank-based security metrics. The proposed model significantly outperforms the bigram baseline across all measures. In particular, the Top-3 success rate increases from $20.56\%$ to $41.87\%$, and the Top-5 success rate increases from $32.41\%$ to $50.38\%$. This implies that, under realistic rate-limited attack scenarios, the probability of successful PIN recovery nearly doubles when using the proposed model. Furthermore, the expected guess rank is reduced from $23.50$ to $17.75$, indicating a more efficient search process. \textit{Notably,} while macro-averaged precision, recall, and F1-score remain low due to class imbalance and large output space (ref. to Table~\ref{tab:missing_digitsPred}), these metrics reflect uniform class-wise performance rather than adversarial effectiveness. In contrast, rank-based metrics (Top-k success rates and expected guess rank) better capture the practical success of an attacker under rate-limited conditions. The observed discrepancy highlights that strong attack performance can coexist with low macro-averaged scores.

\smallskip

Fig.~\ref{fig:comp_prediction} reports accuracy for Decision Tree, Random Forest, Naive Bayes, and the proposed method across one-, two-, and three-digit leakage scenarios.
Also, we observe a consistent performance trend between the proposed context-conditioned model and the HMM baseline across all missing-digit scenarios (evaluated on a representative subset of 50{,}000 PINs). While the HMM achieves comparable accuracy in some cases, the proposed model consistently outperforms it in terms of precision, recall, and F1-score, indicating stronger class-wise predictive reliability.
For example, in the case of single-digit inference, the proposed model yields substantially higher F1-scores (e.g., $\sim$25.7\% vs.\ $\sim$11.4\% for $d_1 \mid d_2d_3d_4$, and $\sim$30.3\% vs.\ $\sim$18.9\% for $d_3 \mid d_1d_2d_4$), along with significantly improved precision. In two-digit prediction tasks, this advantage becomes more pronounced, with F1-scores typically $2$--$4\times$ higher. Under three-digit missing scenarios, although both models degrade due to increased uncertainty, the proposed model maintains consistently higher F1-scores and recall, reflecting more stable inference behavior.

These results, together, suggest that, compared to the HMM and other classical ML baselines, the proposed context-conditioned probabilistic model more effectively captures cross-position dependencies and avoids bias toward frequent patterns, resulting in improved performance across varying levels of prediction difficulty.

\section{Conclusion and Future Work}
\label{sec:conclusionFW}
We model PIN entry as a noisy human-machine communication channel whose reliability degrades under partial information leakage, and analyze this effect using a context-conditioned probabilistic inference framework. The proposed model leverages positional context to predict missing digits under partial observation using smoothed conditional distributions and fallback priors.
Our results show that even limited leakage significantly increases adversarial success, challenging the conventional binary notion of secure versus compromised authentication. While classical baselines, including tree-based models and HMMs, achieve reasonable performance, the proposed approach consistently improves precision, recall, and F1-score across all missing-digit scenarios. Notably, it maintains meaningful predictive accuracy even under severe leakage conditions (e.g., 12.12\% for three missing digits), indicating persistent structural patterns in user-selected PINs.
These findings suggest that partial compromise is not purely incremental, but can be systematically exploited, and that modeling cross-position dependencies is important for realistic security assessment in human-in-the-loop authentication systems.
Future work will focus on (i) evaluating end-to-end IoT authentication scenarios with physical-layer signal acquisition, (ii) expanding datasets to capture cultural and temporal variations in PIN selection, and (iii) exploring more advanced probabilistic and learning-based models for improved generalization under heterogeneous leakage conditions.

\section*{Acknowldgement}
This work is supported in part by the Natural Sciences and Engineering Research Council of Canada under the Discovery and CREATE TRAVERSAL programs.

\begin{comment}
\newpage 
%Put this in Section IV (Results) Right after: Table II or HMM vs baseline comparison
\textcolor{green}{A consistent performance trend is observed between the proposed context-conditioned model and the standard HMM baseline across all missing-digit scenarios (evaluated on a representative subset of 50{,}000 PINs). While the HMM achieves comparable accuracy in some cases, the proposed model consistently outperforms it in terms of precision, recall, and F1-score, indicating stronger class-wise predictive reliability.\\
For single-digit inference, the proposed model yields substantially higher F1-scores (e.g., $\sim$25.7\% vs.\ $\sim$11.4\% for $d_1 \mid d_2d_3d_4$, and $\sim$30.3\% vs.\ $\sim$18.9\% for $d_3 \mid d_1d_2d_4$), along with significantly improved precision. In two-digit prediction tasks, this advantage becomes more pronounced, with F1-scores typically $2$--$4\times$ higher. Under three-digit missing scenarios, although both models degrade due to increased uncertainty, the proposed model maintains consistently higher F1-scores and recall, reflecting more stable inference behavior.\\
These results suggest that, compared to the HMM baseline, the proposed context-conditioned probabilistic model more effectively captures cross-position dependencies and avoids bias toward frequent patterns, resulting in improved performance across varying levels of prediction difficulty.}
\end{comment}

\balance 
\bibliographystyle{IEEEtran}
%\bibliography{reference}

\begin{thebibliography}{10}
\providecommand{\url}[1]{#1}
\csname url@samestyle\endcsname
\providecommand{\newblock}{\relax}
\providecommand{\bibinfo}[2]{#2}
\providecommand{\BIBentrySTDinterwordspacing}{\spaceskip=0pt\relax}
\providecommand{\BIBentryALTinterwordstretchfactor}{4}
\providecommand{\BIBentryALTinterwordspacing}{\spaceskip=\fontdimen2\font plus
\BIBentryALTinterwordstretchfactor\fontdimen3\font minus \fontdimen4\font\relax}
\providecommand{\BIBforeignlanguage}[2]{{%
\expandafter\ifx\csname l@#1\endcsname\relax
\typeout{** WARNING: IEEEtran.bst: No hyphenation pattern has been}%
\typeout{** loaded for the language `#1'. Using the pattern for}%
\typeout{** the default language instead.}%
\else
\language=\csname l@#1\endcsname
\fi
#2}}
\providecommand{\BIBdecl}{\relax}
\BIBdecl

\bibitem{singh2024reliability}
K.~Singh, M.~Yadav, Y.~Singh, D.~Barak, A.~Saini, and F.~Moreira, ``{Reliability on the Internet of Things with Designing Approach for Exploratory Analysis},'' \emph{Frontiers in Computer Science}, vol.~6, p. 1382347, 2024.

\bibitem{lee2024minding}
E.-J. Lee, ``{Minding the source: Toward an Integrative Theory of Human-Machine Communication},'' \emph{Human Communication Research}, vol.~50, no.~2, pp. 184--193, 2024.

\bibitem{reliability-communication2015}
M.~Mushi, E.~Murphy-Hill, and R.~Dutta, ``The human factor: A challenge for network reliability design,'' in \emph{Intl Conf. on the Design of Reliable Communication Networks}, 2015, pp. 115--118.

\bibitem{modern_shoulderSurfing}
O.~Wiese and V.~Roth, ``{See You Next Time: A Model for Modern Shoulder Surfers},'' in \emph{Proceedings of the 18th International Conference on Human-Computer Interaction with Mobile Devices and Services}, ser. MobileHCI '16.\hskip 1em plus 0.5em minus 0.4em\relax New York, NY, USA: Association for Computing Machinery, 2016, p. 453–464.

\bibitem{bace2022privacyscout}
M.~B{\^a}ce, A.~Saad, M.~Khamis, S.~Schneegass, and A.~Bulling, ``{PrivacyScout: Assessing Vulnerability to Shoulder Surfing on Mobile Devices},'' \emph{Proceedings on Privacy Enhancing Technologies}, 2022.

\bibitem{tan2023hollow}
J.~Tan and D.~K. Sarmah, ``{Hollow-Pass: A Dual-View Pattern Password Against Shoulder-Surfing Attacks},'' in \emph{Intl Symp. on Cyber Security, Cryptology, and Machine Learning}.\hskip 1em plus 0.5em minus 0.4em\relax Springer, 2023, pp. 251--272.

\bibitem{hu2023password}
J.~Hu, H.~Wang, T.~Zheng, J.~Hu, Z.~Chen, H.~Jiang, and J.~Luo, ``{Password-stealing without Hacking: Wi-Fi Enabled Practical Keystroke Eavesdropping},'' in \emph{Proceedings of the 2023 ACM SIGSAC conference on computer and communications security}, 2023, pp. 239--252.

\bibitem{kwon-shoulderSurfing}
T.~Kwon, S.~Shin, and S.~Na, ``{Covert Attentional Shoulder Surfing: Human Adversaries Are More Powerful Than Expected},'' \emph{IEEE Transactions on Systems, Man, and Cybernetics: Systems}, vol.~44, no.~6, pp. 716--727, 2014.

\bibitem{chakraborty2025your}
N.~Chakraborty and M.~Zulkernine, ``{Is Your PIN Safe Against Advanced Human-Centric Shoulder Surfing?}'' in \emph{IEEE 49th Annual Computers, Software, and Applications Conf.}\hskip 1em plus 0.5em minus 0.4em\relax IEEE, 2025, pp. 2307--2312.

\bibitem{nileshIoT-PIN}
N.~Chakraborty, J.-Q. Li, S.~Mondal, C.~Luo, H.~Wang, M.~Alazab, F.~Chen, and Y.~Pan, ``{On Designing a Lesser Obtrusive Authentication Protocol to Prevent Machine-Learning-Based Threats in Internet of Things},'' \emph{IEEE Internet of Things J.}, vol.~8, no.~5, pp. 3255--3267, 2021.

\bibitem{IoTSec-PIN2018}
X.~Su, Z.~Wang, X.~Liu, C.~Choi, and D.~Choi, ``{Study to Improve Security for IoT Smart Device Controller: Drawbacks and Countermeasures},'' \emph{Security and Communication Networks}, vol. 2018, no.~1, p. 4296934, 2018.

\bibitem{chakraborty2014improved}
N.~Chakraborty and S.~Mondal, ``{An Improved Methodology Towards Providing Immunity Against Weak Shoulder Surfing Attack},'' in \emph{Information Systems Security: 10th International Conference, ICISS 2014, Hyderabad, India, December 16-20, 2014, Proceedings 10}.\hskip 1em plus 0.5em minus 0.4em\relax Springer, 2014, pp. 298--317.

\bibitem{roth2004pin}
V.~Roth, K.~Richter, and R.~Freidinger, ``{A PIN-entry Method Resilient Against Shoulder Surfing},'' in \emph{Proceedings of the 11th ACM Conference on Computer and Communications Security}, 2004, pp. 236--245.

\bibitem{skullsecurity-passwords}
R.~Bowes, ``{SkullSecurity Password Datasets},'' \url{https://www.skullsecurity.org/wiki/Passwords}, 2010, accessed: 2025-10-30.

\bibitem{kokila2025authentication}
M.~Kokila and S.~Reddy, ``Authentication, access control and scalability models in internet of things security--a review,'' \emph{Cyber Security and Applications}, vol.~3, 2025.

\bibitem{schneegass2022investigation}
S.~Schneegass, A.~Saad, R.~Heger, S.~Delgado~Rodriguez, R.~Poguntke, and F.~Alt, ``{An Investigation of Shoulder Surfing Attacks on Touch-Based Unlock Events},'' \emph{Proceedings of the ACM on Human-Computer Interaction}, vol.~6, no. MHCI, pp. 1--14, 2022.

\bibitem{shukla2019stealing}
D.~Shukla and V.~V. Phoha, ``{Stealing Passwords by Observing Hands Movement},'' \emph{IEEE Transactions on Information Forensics and Security}, vol.~14, no.~12, pp. 3086--3101, 2019.

\bibitem{wodo2016thermal}
W.~Wodo and L.~Hanzlik, ``{Thermal Imaging Attacks on Keypad Security Systems},'' in \emph{SECRYPT}, 2016, pp. 458--464.

\bibitem{varma2022vibroauth}
M.~Varma, S.~Watson, L.~Chan, and R.~Peiris, ``{VibroAuth: Authentication with Haptics based Non-Visual, Rearranged Keypads to Mitigate Shoulder Surfing Attacks},'' in \emph{International Conference on Human-Computer Interaction}.\hskip 1em plus 0.5em minus 0.4em\relax Springer, 2022, pp. 280--303.

\bibitem{ma2014study}
J.~Ma, W.~Yang, M.~Luo, and N.~Li, ``{A Study of Probabilistic Password Models},'' in \emph{2014 IEEE Symposium on Security and Privacy}.\hskip 1em plus 0.5em minus 0.4em\relax IEEE, 2014, pp. 689--704.

\bibitem{thai2024study}
B.~L.~T. Thai and H.~Tanaka, ``{A Study on Markov-Based Password Strength Meters},'' \emph{IEEE Access}, vol.~12, pp. 69\,066--69\,075, 2024.

\bibitem{weir2009password}
M.~Weir, S.~Aggarwal, B.~De~Medeiros, and B.~Glodek, ``{Password Cracking Using Probabilistic Context-Free Grammars},'' in \emph{2009 30th IEEE symposium on security and privacy}.\hskip 1em plus 0.5em minus 0.4em\relax IEEE, 2009, pp. 391--405.

\bibitem{wang2017understanding}
D.~Wang, Q.~Gu, X.~Huang, and P.~Wang, ``{Understanding Human-Chosen PINs: Characteristics, Distribution and Security},'' in \emph{Proc. of the ACM on Asia Conf. on Computer and Communications Security}, 2017, pp. 372--385.

\end{thebibliography}

\end{document}